\documentclass[12pt]{article}
\usepackage{amsmath}
\usepackage{amssymb}
\usepackage{amsfonts}

\renewcommand{\theequation}{\arabic{section}.\arabic{equation}}
\DeclareMathOperator{\csch}{csch}

\title{Deformed Shape Invariant Superpotentials in Quantum Mechanics and Expansions in Powers of $\hbar$}
\author{Christiane Quesne\\ 
{\small\sl Physique Nucl\'eaire Th\'eorique et Physique Math\'ematique,  Universit\'e Libre de Bruxelles,} \\ 
{\small\sl Campus de la Plaine CP229, Boulevard~du Triomphe, B-1050 Brussels, Belgium} \\
{\small\sl E-Mail: cquesne@ulb.ac.be}}
\date{ }
\begin{document}
\baselineskip=22pt plus 1pt minus 1pt
\maketitle
\begin{abstract}
We show that the method developed by Gangopadhyaya, Mallow, and their coworkers to deal with (translationally) shape invariant potentials in supersymmetric quantum mechanics and consisting in replacing the shape invariance condition, which is a difference-differential equation, by an infinite set of partial differential equations can be generalized to deformed shape invariant potentials in deformed supersymmetric quantum mechanics. The extended method is illustrated by several examples, corresponding both to $\hbar$-independent superpotentials and to a superpotential explicitly depending on $\hbar$. 
\end{abstract}


%
%
\newpage
\section{Introduction}

Exactly solvable (ES) Schr\"odinger equations (SE) allow to understand some physical phenomena and to test some approximation schemes. Supersymmetric quantum mechanics (SUSYQM) \cite{cooper, junker, bagchi00, gango10} is known to be a very powerful method for generating such ES models, especially whenever the corresponding potential is (translationally) shape invariant (SI) \cite{genden}. SUSYQM may be considered as a modern version of the old Darboux transformation \cite{darboux} and of the factorization method used by Schr\"odinger \cite{schro40, schro41a, schro41b} and by Infeld and Hull \cite{infeld51}.\par
%
%
Some ten years ago, the list of (translationally) SI potentials, whose bound-state wavefunctions can be expressed in terms of classical orthogonal polynomials (COP) \cite{cooper} has been completed (see, e.g., \cite{gomez14} and references quoted therein) by introducing \cite{cq08, bagchi09, cq09a, odake} some rational extensions of these potentials, connected with the novel field of exceptional orthogonal polynomials (EOP) \cite{gomez09}. The latter are polynomial sets which are orthogonal and complete, but, in contrast with COP, admit a finite number of gaps in the sequence of their degrees.\par
%
%
ES models for some unconventional SE are also very interesting. These unconventional equations may be of three different kinds. They may occur whenever the standard commutation relations are replaced by deformed ones, associated with nonzero minimal uncertainties in position and/or momentum \cite{kempf94, hinrichsen, kempf97}, as suggested by several investigations in string theory and quantum gravity \cite{witten}. They may also appear whenever the constant mass of the conventional SE is replaced by a position-dependent mass (PDM). The latter is an essential ingredient in the study of electronic properties of semiconductor heterostructures \cite{bastard, weisbuch}, quantum wells and quantum dots \cite{serra, harrison}, helium clusters \cite{barranco}, graded crystals \cite{geller}, quantum liquids \cite{arias}, metal clusters \cite{puente}, nuclei \cite{ring, bonatsos}, nanowire structures \cite{willatzen}, and neutron stars \cite{chamel}. A third possibility corresponds to the replacement of the Euclidean space by a curved one. The study of the Kepler-Coulomb problem on the sphere dates back to the work of Schr\"odinger \cite{schro40}, Infeld \cite{infeld41}, Stevenson \cite{stevenson}, and was generalized to a hyperbolic space by Infeld and Schild \cite{infeld45}. Since then, many studies have been devoted to this topic (see, e.g., \cite{kalnins96, kalnins97}).\par
%
%
As shown elsewhere \cite{cq04}, there are some intimate connections between these three types of unconventional SE, occurring whenever a specific relation exists between the deforming function $f(x)$, the PDM $m(x)$, and the (diagonal) metric tensor $g(x)$. Such unconventional SE may then be discussed in the framework of deformed SUSYQM (DSUSYQM), where the standard SI condition is replaced by a deformed one (DSI) \cite{bagchi05, cq09b}. On starting from the known superpotentials  of SI potentials \cite{cooper}, a procedure has been devised in \cite{bagchi05} to maintain the solvability of the DSI condition, thereby resulting in a list of deformed superpotentials and deforming functions giving rise to bound-state spectra. In such deformed case, physically acceptable wavefunctions have not only to be square integrable on the defining interval of the potential, but also must ensure the Hermiticity of the Hamiltonian. More recently \cite{cq16}, this list of deformed superpotentials and deforming functions has been completed by considering the case of some rationally-extended potentials, connected with one-indexed families of EOP.\par
%
%
In the case of conventional SUSYQM, Gangopadhyaya, Mallow, and their coworkers proposed an interesting approach to SI potentials, consisting in replacing the SI condition, which is a difference-differential equation, by an infinite set of partial differential equations. The latter is obtained by expanding the superpotential in powers of $\hbar$ and expressing that the coefficient of each power must separately vanish \cite{gango08}. This procedure enabled them to prove that the SI superpotentials connected with COP are those with no explicit dependence on $\hbar$, while the new ones related to EOP have such an explicit dependence \cite{bougie10}. They also showed that the list of the former given in \cite{cooper} is complete \cite{bougie12} and constructed a novel example of SI superpotential with an explicit $\hbar$-dependence \cite{bougie15}. Furthermore, they encountered a pathway for going from those superpotentials of \cite{cooper} corresponding to SE that can be reduced to the confluent hypergeometric equation to those related to SE connected with the hypergeometric equation \cite{mallow}.\par
%
%
It is the purpose of the present paper to propose an extension of the approach of Gangopadhyaya, Mallow, and their coworkers to the case of DSI potentials in DSUSYQM, both without and with explicit dependence of the superpotential on $\hbar$. We plan to illustrate this method by re-examining the known pairs of deformed superpotentials and deforming functions of \cite{bagchi05, cq09b, cq16} along these lines. In the present work, we restrict ourselves to unbroken DSUSYQM and only consider the discrete part of the spectrum.\par
%
%
After reviewing the general formalism of DSUSYQM and obtaining the DSI condition in Section~2, we will show in Section~3 how to convert such a condition into a set of partial differential equations in the case where the superpotential does not contain any dependence on $\hbar$. The case where the superpotential has such an explicit dependence is then treated in Section~4. Finally, Section~5 contains the conclusion.\par
%
%
\section{Deformed Shape Invariance in Deformed Supersymmetric Quantum Mechanics}

In DSUSYQM \cite{bagchi05, cq09b, cq16}, a general Hamiltonian $H_-$ is written in terms of linear operators
\begin{equation}
  A^{\pm} = A^{\pm}(a) = \mp \hbar \sqrt{f(x)} \frac{d}{dx} \sqrt{f(x)} + W(x,a),  \label{eq:A}
\end{equation}
where $f(x)$ is some positive-definite function of $x$, known as the deforming function, and $W(x,a)$ is a real function of $x$ and a parameter $a$, called the superpotential. Both $f(x)$ and $W(x,a)$ in general depend on some extra parameters. The Hamiltonian $H_-$ is given by
\begin{equation}
  H_- = A^+ A^- = - \hbar^2 \sqrt{f(x)} \frac{d}{dx} f(x) \frac{d}{dx} \sqrt{f(x)} + V_-(x,a),  \label{eq:H-}
\end{equation}
where
\begin{equation}
  V_-(x,a) = W^2(x,a) - \hbar f(x) \frac{dW(x,a)}{dx}.  \label{eq:V-}
\end{equation}
It may be interpreted \cite{cq04} as a Hamiltonian describing a PDM system with $m(x) = 1/f^2(x)$, the ordering of the latter and the differential operator $d/dx$ being that proposed by Mustafa and Mazharimousavi \cite{mustafa}, or as a Hamiltonian in a curved space with a diagonal metric tensor $g(x) = 1/f^2(x)$.\par
%
%
The product of operators $A^- A^+$ generates the so-called partner of $H_-$,
\begin{equation}
  H_+ = A^- A^+ = - \hbar^2 \sqrt{f(x)} \frac{d}{dx} f(x) \frac{d}{dx} \sqrt{f(x)} + V_+(x,a),
\end{equation}
with
\begin{equation}
  V_+(x,a) = W^2(x,a) + \hbar f(x) \frac{dW(x,a)}{dx}.  \label{eq:V+}
\end{equation}
The pair of Hamiltonians intertwine with $A^+$ and $A^-$ as
\begin{equation}
  A^- H_- = H_+ A^-, \qquad A^+ H_+ = H_- A^+.  \label{eq:intertwine}
\end{equation}
\par
%
%
The Hamiltonian $H_-$ is assumed to have a ground-state wavefunction $\psi_0^{(-)}(x,a)$, such that
\begin{equation}
  A^- \psi_0^{(-)}(x,a) = 0.
\end{equation}
From (\ref{eq:A}) and (\ref{eq:H-}), the latter is therefore such that $E_0^{(-)} = 0$ and
\begin{equation}
  \psi_0^{(-)}(x,a) = \frac{N_0^{(-)}}{\sqrt{f(x)}} \exp {\left(- \int^x \frac{W(x',a)}{\hbar f(x')} dx'\right)},
  \label{eq:gswf}
\end{equation}
where $N_0^{(-)}$ is sone normalization coefficient.\par
%
%
The intertwining relations (\ref{eq:intertwine}) then imply the following isospectrality relationships among the eigenvalues and eigenfunctions of the two partners,
\begin{align}
  & E_{n+1}^{(-)} = E_n^{(+)}, \\
  & \psi_n^{(+)}(x,a) = \frac{A^-}{\sqrt{E_n^{(+)}}} \psi_{n+1}^{(-)}(x,a), \qquad 
       \psi_{n+1}^{(-)}(x,a) = \frac{A^+}{\sqrt{E_n^{(+)}}} \psi_n^{(+)}(x,a),
\end{align}
for all $n \ge 0$ such that physically acceptable wavefunctions exist. In the deformed case considered here, this imposes that they satisfy two conditions \cite{bagchi05}:\par
\noindent (i) As for conventional SE, they should be square integrable on
the (finite or infinite) interval of definition $(x_1,x_2)$ of the potentials $V_{\pm}(x,a)$, i.e.,
\begin{equation}
  \int_{x_1}^{x_2} dx\,  |\psi_n^{(\pm)}(x,a)|^2 < \infty.  \label{eq:bound-1} 
\end{equation}
\noindent (ii) They should ensure the Hermiticity of $H_{\pm}$. This amounts to the condition 
\begin{equation}
  |\psi_n^{(\pm)}(x,a))|^2 f(x) \to 0 \qquad \text{for $x \to x_1$ and $x \to x_2$},  \label{eq:bound-2}  
\end{equation}
which implies an additional restriction whenever $f(x) \to \infty$ for $x \to x_1$ and/or $x \to x_2$. Equations (\ref{eq:bound-1}) and (\ref{eq:bound-2}) ensure the self-adjointness of $H_{\pm}$ and guarantee the relation $(A^{\pm})^{\dagger} = A^{\mp}$.\par
%
%
The knowledge of the eigenvalues and eigenfunctions of $H_-$ automatically implies the same for its partner $H_+$ (or vive versa). However, whenever the partner potentials $V_-$ and $V_+$ are similar in shape and differ only in the parameters that appear in them, i.e.,
\begin{equation}
  V_+(x,a_0) + g(a_0) = V_-(x,a_1) + g(a_1),  \label{eq:SI}
\end{equation}
where $a_1$ is some function of $a_0$ and $g(a_0)$, $g(a_1)$ do not depend on $x$, then the spectrum of either Hamiltonian can be derived without reference to its partner. Here we restrict ourselves to the case of translational (or additive) shape invariance, i.e., $a_1$ and $a_0$ only differ by some additive constant. Considering then a set of parameters $a_i$, $i=0, 1, 2, \ldots$, and extending condition (\ref{eq:SI}) to 
\begin{equation}
  V_+(x,a_i) + g(a_i) = V_-(x,a_{i+1}) + g(a_{i+1}), \qquad i=0, 1, 2, \ldots,
\end{equation}
we get from Eqs.~(\ref{eq:V-}) and (\ref{eq:V+}) the so-called DSI condition
\begin{equation}
  W^2(x,a_i) + \hbar f(x) \frac{dW(x,a_i)}{dx} + g(a_i) = W^2(x,a_{i+1}) - \hbar f(x) \frac{dW(x,a_{i+1})}{dx} 
  + g(a_{i+1}).  \label{eq:DSI}
\end{equation}
The eigenvalues and eigenfunctions of $H_-$ turn out to be given by
\begin{align}
  & E_n^{(-)}(a_0) = g(a_n) - g(a_0), \qquad n=0, 1, 2, \ldots, \label{eq:energy} \\
  & \psi_n^{(-)}(x,a_0) \propto A^+(a_0) A^+(a_1) \ldots A^+(a_{n-1}) \psi_0^{(-)}(x,a_n), \qquad n=0, 1,
      2, \ldots,
\end{align}
with $\psi_0^{(-)}(x,a_n)$ as expressed in (\ref{eq:gswf}).\par
%
%
\section{\boldmath Deformed Shape Invariance for Superpotentials with no Explicit Dependence on $\hbar$}

\setcounter{equation}{0}

As in \cite{gango08, bougie10, bougie12, bougie15, mallow}, let us assume that the additive constant that allows to get $a_{i+1}$ from $a_i$ is just $\hbar$, i.e., $a_{i+1} = a_i + \hbar$. Note that with respect to conventions used elsewhere where the system of units is such that $\hbar=1$, this implies some parameter renormalization. In the appendix, we summarize the transformations that have to be carried out on the parameters and possibly the variable used in \cite{bagchi05, cq09b, cq16} in order to arrive at the conventions employed here.\par
%
%
In the present section, we will also suppose that the dependence of $W(x,a_i)$ on $\hbar$ is entirely contained in $a_i$, thus leaving the case of an explicit dependence of $W$ on $\hbar$ to Section~4.\par
%
%
Since Eq.~(\ref{eq:DSI}) must hold for an arbitrary value of $\hbar$, we can expand it in powers of $\hbar$ and require that the coefficient of each power vanishes. It is straightforward to show that the coefficient of $\hbar$ leads to the condition
\begin{equation}
  W \frac{\partial W}{\partial a} - f(x) \frac{\partial W}{\partial x} + \frac{1}{2} \frac{dg}{da} = 0.
  \label{eq:condition1}
\end{equation}
Then, the coefficient of $\hbar^2$ yields
\begin{equation}
  \frac{\partial}{\partial a} \left[W \frac{\partial W}{\partial a} - f(x) \frac{\partial W}{\partial x} + \frac{1}{2}
  \frac{dg}{da}\right] = 0,
\end{equation}
which is automatically satisfied if Eq.~(\ref{eq:condition1}) is so. Finally, the coefficient of $\hbar^n$ for $n \ge 3$ gives the condition
\begin{equation}
  \frac{(2-n) f(x)}{n!} \frac{\partial^n W}{\partial a^{n-1} \partial x} = 0, \qquad n=3, 4, \ldots.
\end{equation}
 This set of equations is satisfied provided
 \begin{equation}
  \frac{\partial^3 W}{\partial a^2 \partial x} = 0.  \label{eq:condition2}
\end{equation}
We are therefore left with two independent conditions (\ref{eq:condition1}) and (\ref{eq:condition2}). This is similar to what happens in SUSYQM \cite{bougie10, bougie12}, the only difference being the appearance of the deforming function $f(x)$ in the first equation.\par
%
%
Before giving the set of results, we shall discuss in detail two examples, a simple one and a more involved one.\par
%
%
\subsection{Example of the P\"oschl-Teller Potential}

Let us consider a deforming function $f(x) = 1 + \alpha \sin^2 x$ with $-1 < \alpha\ne 0$ and $-\frac{\pi}{2} < x < \frac{\pi}{2}$, as well as a superpotential
\begin{equation}
  W(x,a) = (1+\alpha)a\tan x, \qquad -\frac{\pi}{2} < x < \frac{\pi}{2}, 
\end{equation}
where
\begin{equation}
  a = \frac{1}{2(1+\alpha)} [(1+\alpha)\hbar + \Delta], \qquad \Delta = \sqrt{(1+\alpha)^2 \hbar^2 
  + 4A(A-\hbar)}, \qquad A>\hbar. 
\end{equation}
We note that this $W$ automatically satisfies Eq.~(\ref{eq:condition2}) and that, on inserting it in Eq.~(\ref{eq:condition1}), we obtain
\begin{equation}
  \frac{dg}{da} = 2(1+\alpha)a,
\end{equation}
from which  $g(a) = (1+\alpha) a^2$, up to some additive constant.\par
%
%
{}From Eq.~(\ref{eq:V-}), the starting potential can be written as
\begin{align}
  V_-(x,a) &= (1+\alpha)^2 a(a-\hbar) \sec^2x  - (1+\alpha)a[(1+\alpha)a - \hbar \alpha] \nonumber \\
  &= A(A-\hbar) \sec^2x - \left\{A(A-\hbar) + \frac{1}{2}\hbar [(1+\alpha)\hbar + \Delta]\right\},
\end{align}
and therefore corresponds to the P\"oschl-Teller potential $V=A(A-\hbar)\sec^2x$ with ground-state energy $E_0 = A(A-\hbar) + \frac{1}{2}\hbar [(1+\alpha)\hbar + \Delta]$. On the other hand, from (\ref{eq:energy}), we get
\begin{equation}
  E^{(-)}_n = g(a+n\hbar) - g(a) = \hbar^2 (1+\alpha) n(n+1) + \hbar \Delta n.
\end{equation}
\par
%
%
The results obtained here may be compared with those derived in \cite{cq09b} for $\bar{V} = \bar{A}(\bar{A}-1) \sec^2\bar{x}$, with bound-state energies $\bar{E}_n = (E^{(-)}_n + E_0)/{\hbar}^2 = (\bar{\lambda}+n)^2 - \bar{\alpha}(\bar{\lambda}-n^2)$, where $\bar{\lambda} = (1+\bar{\alpha})a/\hbar$ is changed into $\bar{\lambda}+1+\bar{\alpha}$ when going to the partner.\par
%
%
\subsection{Example of the Radial Harmonic Oscillator Potential}

Let us now consider a deforming function $f(x) = 1 + \alpha x^2$ with $\alpha>0$ and $0<x<\infty$, as well as a superpotential
\begin{equation}
  W(x,a) = a\left(- \frac{1}{x} + \alpha x\right) - b \left(\frac{1}{x} + \alpha x\right), \qquad 0<x<\infty,
\end{equation}
where
\begin{align}
  a &= \frac{1}{2} \left(l + 1 + \frac{1}{2}\hbar + \frac{\Delta}{2\alpha}\right), \qquad
  b = \frac{1}{2} \left(l + 1 - \frac{1}{2}\hbar - \frac{\Delta}{2\alpha}\right), \nonumber \\ 
  \Delta &= \sqrt{\omega^2 + \hbar^2 \alpha^2}, \qquad \omega>0, \qquad l=0,1,2,\ldots.
\end{align}
Here, when going to the partner $a$ is assumed to change into $a+\hbar$, while $b$ remains constant. This $W$ automatically satisfies Eq.~(\ref{eq:condition2}) again and Eq.~(\ref{eq:condition1}) leads to
\begin{equation}
  \frac{dg}{da} = 8\alpha a,
\end{equation}
from which $g(a) = 4\alpha a^2$, up to some additive constant.\par
%
%
Equation~(\ref{eq:V-}) shows that the starting potential is given by 
\begin{align}
  V_-(x,a) &= \alpha^2 (a-b) (a-b-\hbar)x^2 + \frac{(a+b)(a+b-\hbar)}{x^2} - 2\alpha(a^2-b^2+\hbar a)
       \nonumber \\
  &= \frac{1}{4}\omega^2 x^2 + \frac{(l+1)(l+1-\hbar)}{x^2} - \left[\left(l+1+\frac{\hbar}{2}\right)\Delta
       + \hbar \alpha\left(2l+2+\frac{1}{2}\hbar\right)\right]
\end{align}
and therefore corresponds to the radial harmonic oscillator (RHO) potential $V = \frac{1}{4}\omega^2 x^2 + \frac{(l+1)(l+1-\hbar)}{x^2}$ with ground-state energy $E_0 = \left(l+1+\frac{\hbar}{2}\right)\Delta
+ \hbar \alpha\left(2l+2+\frac{1}{2}\hbar\right)$. Furthermore, Eq.~(\ref{eq:energy}) leads to 
\begin{equation}
  E^{(-)}_n = g(a+n\hbar) - g(a) = 4\hbar\alpha n\left[\left(n + \frac{1}{2}\right)\hbar + l +1\right]
  + 2\hbar n\Delta.
\end{equation}
\par
%
%
These results are comparable with those obtained in \cite{cq09b} for $\bar{V} = \frac{1}{4}\bar{\omega}^2 \bar{x}^2 + \bar{l}(\bar{l}+1)/\bar{x}^2$ with bound-state energies $\bar{E}_n = E^{(-)}_n + E_0 = 2\bar{\lambda}\bar{\mu} - \bar{\alpha}\bar{\lambda} + \bar{\mu} - 4(\bar{\alpha}\bar{\lambda}-\bar{\mu})n + 4\bar{\alpha}n^2$, where $\bar{\lambda} = - (a+b)/\hbar$ and $\bar{\mu} = (a-b)\hbar\alpha$ are changed into $\bar{\lambda}-1$ and $\bar{\mu}+\bar{\alpha}$ when going to the partner, respectively.\par
%
%
\subsection{Lists of Results}
On proceeding as in Subsections 3.1 and 3.2, we have analyzed the other sets of potentials and deforming functions considered in \cite{bagchi05, cq09b, cq16}. They include the Scarf I (S), radial Coulomb (C), Morse (M), Eckart (E), Rosen-Morse I (RM), shifted harmonic oscillator (SHO), deformed radial harmonic oscillator (DRHO), and deformed radial Coulomb (DC) potentials. The list of them is given in Table~1 in the notations used in this paper. In all the cases, except for the PT and DC potentials, the deformed superpotential is written in terms of two combinations of parameters, the  first one $a$ being changed into $a+\hbar$ and the second one $b$ remaining constant when going to the partner. The corresponding results are listed in Table~2. In all cases, it turns out that Eq.~(\ref{eq:condition2}) is automatically satisfied, while Eq.~(\ref{eq:condition1}) leads to the expressions of $g(a)$ listed in Table~3, together with the resulting energies $E^{(-)}_n$.\par
%
%
\begin{table}[p]

\caption{Potentials and deforming functions.}

\begin{center}
\begin{tabular}{lll}
  \hline\\[-0.2cm]
  Name & $V$ & $f$ \\[0.2cm]
  \hline\\[-0.2cm]
  PT & $A(A-\hbar) \sec^2x$ & $1+\alpha\sin^2x$ \\[0.2cm]
  & $-\frac{\pi}{2}<x<\frac{\pi}{2}$, $A>\hbar$ & $-1< \alpha\ne 0$ \\[0.2cm]
  RHO & $\frac{1}{4}\omega^2 x^2 + \frac{(l+1)(l+1-\hbar)}{x^2}$ & $1+\alpha x^2$ \\[0.2cm]
  & $0<x<+\infty$ & $\alpha>0$ \\[0.2cm]
  S & $[A(A-\hbar)+B^2]\sec^2x - B(2A-\hbar)\sec x \tan x $ & $1+\alpha\sin x$ \\[0.2cm]
  & $-\frac{\pi}{2}<x<\frac{\pi}{2}$, $A-\hbar>B>0$ & $0<|\alpha|<1$ \\[0.2cm]
  C & $-\frac{e^2}{x} + \frac{(l+1)(l+1-\hbar)}{x^2}$ & $1+\alpha x$ \\[0.2cm]
  & $0<x<+\infty$ & $\alpha>0$ \\[0.2cm]
  M & $B^2 e^{-2x} - B(2A+\hbar) e^{-x}$ & $1+\alpha e^{-x}$ \\[0.2cm]
  & $-\infty < x < +\infty$, $A, B>0$ & $\alpha>0$ \\[0.2cm]
  E & $A(A-\hbar) \csch^2 x - 2B \coth x$ & $1+\alpha e^{-x} \sinh x$ \\[0.2cm]
  & $0<x<+\infty$, $A \ge \frac{3}{2}\hbar$, $B>A^2$ & $-2 < \alpha\ne 0$ \\[0.2cm]
  RM & $A(A-\hbar) \csc^2 x + 2B\cot x$ & $1+\sin x(\alpha \cos x + \beta \sin x)$ \\[0.2cm]
  & $0<x<\pi$, $A\ge \frac{3}{2}\hbar$ & $\frac{|\alpha|}{2} < \sqrt{1+\beta}$, $\beta>-1$ \\[0.2cm]
  SHO & $\frac{1}{4}\omega^2 \left(x-\frac{2d}{\omega}\right)^2$ & $1 + \alpha x^2 + 2\beta x$ \\[0.2cm]
  & $-\infty<x<+\infty$ & $\alpha > \beta^2 \ge 0$ \\[0.2cm]
  DRHO & $\frac{\omega(\omega+2\hbar\lambda)x^2}{4(1+\lambda x^2)} + \frac{(l+1)(l+1-\hbar)}{x^2}$ &
        $\sqrt{1+\lambda x^2}$ \\[0.2cm]
  & $0<x<+\infty$ if $\lambda>0$ & \\[0.2cm]
  & $0<x<1/\sqrt{|\lambda|}$ if $\lambda<0$ \\[0.2cm]
  DC & $- \frac{e^2}{x}\sqrt{1+\lambda x^2} + \frac{(l+1)(l+1-\hbar)}{x^2}$ & $\sqrt{1+\lambda x^2}$ 
       \\[0.2cm]
  & $0<x<+\infty$ if $\lambda>0$ & \\[0.2cm]
  & $0<x<1/\sqrt{|\lambda|}$ if $\lambda<0$ \\[0.2cm]
  \hline 
\end{tabular}
\end{center}

\end{table}
\par
%
%
\begin{table}[p]

\caption{Superpotentials and combinations of parameters.}

\begin{center}
\begin{tabular}{lll}
  \hline\\[-0.2cm]
  Name & $W$ & Parameters \\[0.2cm]
  \hline\\[-0.2cm]
  PT & $(1+\alpha)a\tan x$ & $a = \frac{1}{2(1+\alpha)} [(1+\alpha)\hbar + \Delta]$ \\[0.2cm]
  & & $\Delta = \sqrt{(1+\alpha)^2 \hbar^2 + 4A(A-\hbar)}$ \\[0.2cm]
  RHO & $a\left(-\frac{1}{x}+\alpha x\right) - b\left(\frac{1}{x}+\alpha x\right)$ & $a = \frac{1}{2}\left(
       l+1 + \frac{1}{2}\hbar + \frac{\Delta}{2\alpha}\right)$ \\[0.2cm]
  & & $b = \frac{1}{2}\left(l+1 - \frac{1}{2}\hbar - \frac{\Delta}{2\alpha}\right)$ \\[0.2cm]
  & & $\Delta = \sqrt{\omega^2 + \hbar^2 \alpha^2}$ \\[0.2cm]
  S & $a(\tan x + \alpha \sec x)  $ & $a = \frac{1}{2}\left(\hbar + \frac{\alpha-1}{2\alpha}\Delta_+
       + \frac{\alpha+1}{2\alpha}\Delta_-\right)$ \\[0.2cm]
  & $+ b(\tan x - \alpha \sec x)$ & $b = \frac{1}{4\alpha}[(\alpha+1)\Delta_+ + (\alpha-1)\Delta_-]$ \\[0.2cm]
  & & $\Delta_{\pm} = \sqrt{\frac{1}{4}\hbar^2(1\mp\alpha)^2 + (A\pm B) (A\pm B - \hbar)}$ \\[0.2cm]
  C & $- \frac{a+b}{x} + \frac{2b}{a+b} - \frac{\alpha}{2}(a+b)$ & $a = -\frac{1}{4}\{e^2 + (l+1)[\alpha
       (l+1-\hbar) - 4]\}$ \\[0.2cm]
  & & $b = \frac{1}{4}[e^2 + \alpha(l+1)(l+1-\hbar)]$ \\[0.2cm]
  M & $-\alpha(a+b)e^{-x} - \frac{1}{2}(a+b)$ & $a = - \frac{1}{4\hbar\alpha} \{B^2 + \alpha[B(2A+\hbar)
       - 2\hbar^2] - 2\hbar\Delta\}$ \\[0.2cm]
  & $+ \frac{2\hbar b}{\alpha(a+b)}$ & $b = \frac{B}{4\hbar\alpha} [B+\alpha(2A+\hbar)]$ \\[0.2cm]
  & & $\Delta = \sqrt{4B^2 + \hbar^2\alpha^2}$ \\[0.2cm]
  E & $- (a+b)\coth x + \frac{2\hbar b}{a+b}$ & $a = \frac{1}{2\hbar}[-B + 2\hbar A - \frac{\alpha}{2}
       A(A-\hbar)]$ \\[0.2cm]
  & $- \frac{\alpha}{2}(a+b)$ & $b = \frac{1}{2\hbar}[B + \frac{\alpha}{2}A(A-\hbar)]$ \\[0.2cm]
  RM & $- (a+b)\cot x + \frac{2\hbar b}{a+b}$ & $a = \frac{1}{2\hbar}[B + 2\hbar A - \frac{\alpha}{2}
       A(A-\hbar)]$ \\[0.2cm]
  & $- \frac{\alpha}{2}(a+b)$ & $b = \frac{1}{2\hbar}[-B + \frac{\alpha}{2}A(A-\hbar)]$ \\[0.2cm]
  SHO & $(a+b)(\alpha x + \beta)$ & $a = \frac{\hbar}{2}\left(-\hbar \frac{\beta}{4\alpha}\omega^2 + 1
       + \frac{\Delta}{\hbar\alpha} - \frac{d}{2}\hbar\omega\right)$ \\[0.2cm]
  & $- \frac{2b}{\hbar^2\alpha(a+b)}$ & $b = \frac{1}{2}\hbar^2\omega \left(\frac{\beta}{4\alpha}\omega
      + \frac{d}{2}\right)$ \\[0.2cm]
  & & $\Delta = \sqrt{\omega^2 + \hbar^2\alpha^2}$ \\[0.2cm]
  DRHO & $a \left(-\frac{1}{x}f - \frac{\lambda x}{f}\right)$ & $a = \frac{1}{2}\left(l+1 - \frac{\omega}
      {2\lambda}\right)$ \\[0.2cm]
  & $+ b \left(-\frac{1}{x}f + \frac{\lambda x}{f}\right)$ & $b = \frac{1}{2}\left(l+1 + \frac{\omega}
      {2\lambda}\right)$ \\[0.2cm]
  DC & $- \frac{a}{x}f + \frac{e^2}{2a}$ & $a=l+1$ \\[0.2cm]
  \hline 
\end{tabular}
\end{center}

\end{table}
\par
%
%
\begin{table}[p]

\caption{Functions $g(a)$ and bound-state energies.}

\begin{center}
\begin{tabular}{lll}
  \hline\\[-0.2cm]
  Name & $g$ & $E^{(-)}_n$ \\[0.2cm]
  \hline\\[-0.2cm]
  PT & $(1+\alpha)a^2$ & $\hbar^2 (1+\alpha) n(n+1) + \hbar\Delta n$ \\[0.2cm]
  RHO & $4\alpha a^2$ & $4\hbar\alpha n\left[\left(n+\frac{1}{2}\right)\hbar+l+1\right] + 2\hbar n\Delta$
       \\[0.2cm]
  S & $(1-\alpha^2)a^2$ & $\hbar^2(1-\alpha^2)n(n+1) + \hbar \left[(1+\alpha)\Delta_+ +(1-\alpha)\Delta_-
       \right]n$ \\[0.2cm]
  & $+2(1+\alpha^2)ab$ & \\[0.2cm]
  C & $- \frac{4b^2}{(a+b)^2}$ & $\hbar n(\hbar n +2l+2)[e^2 - \hbar\alpha(l+1)(n+1)]$ \\[0.2cm]
  & $- \frac{\alpha^2}{4}a(a+2b)$ & $ \times [e^2 + \alpha(l+1)(\hbar n +2l+2-\hbar)]$ \\[0.2cm]
  & & $\times [4(l+1)^2(l+1+n\hbar)^2]^{-1}$ \\[0.2cm]
  M & $- \frac{4\hbar^2b^2}{\alpha^2(a+b)^2}$ & $ \hbar n [\Delta +\hbar\alpha(n+1)] [2B(2A+\hbar)
        - \hbar(\Delta+\alpha\hbar)(n+1)]$ \\[0.2cm]
  & $-\frac{1}{4}a(a+2b)$ & $\times [4B^2+2\alpha B(2A+\hbar) + \hbar\alpha(\Delta+\hbar\alpha)(n+1)]$
        \\[0.2cm]
  & & $\times (\Delta+\hbar\alpha)^{-2} [\Delta+(2n+1)\hbar\alpha]^{-2}$ \\[0.2cm]
  E & $- \frac{4\hbar^2b^2}{(a+b)^2}$ & $\hbar n(2A+n\hbar)[4A^2(A+n\hbar)^2]^{-1}$ \\[0.2cm]
  & $- \frac{1}{4}(\alpha+2)^2a(a+2b)$ & $\times\{2B+2A(A+n\hbar)+\alpha A[2A+(n-1)\hbar]\}$ \\[0.2cm]
  & & $\times [2B-2A(A+n\hbar)-\hbar\alpha A(n+1)]$ \\[0.2cm]
  RM & $- \frac{4\hbar^2b^2}{(a+b)^2}$ & $\hbar n(2A+n\hbar)[4A^2(A+n\hbar)^2]^{-1}$ \\[0.2cm]
  & $-\frac{1}{4}(\alpha^2-4\beta-4)a(a+2b)$ & $\times \{4A^2(A+n\hbar)^2+4B^2-4\alpha BA(A-\hbar)$
        \\[0.2cm]
  & & $-\hbar\alpha^2A^2[2A-\hbar+n(2A+n\hbar)] - 4\beta A^2(A+n\hbar)^2\}$ \\[0.2cm]
  SHO & $-\frac{4b^2}{\hbar^4\alpha^2(a+b)^2}$ & $4n[\Delta+(n+1)\hbar\alpha]$ \\[0.2cm]
  & $+(\alpha-\beta^2)a(a+2b)$ & $ \times \{(\Delta+\hbar\alpha)[\Delta+\hbar\alpha(2n+1)]\}^{-2}$ 
        \\[0.2cm]
  & & $ \times \{\hbar^3\alpha(\alpha-\beta^2)(n+1)[2\hbar^2\alpha^2(n+1)+\omega^2(n+2)]$ \\[0.2cm]
  & & $ +d\omega^2(\beta+\hbar\alpha d) + \frac{1}{4}\hbar\omega^4$ \\[0.2cm]
  & & $ +\hbar^2\Delta(\alpha-\beta^2)(n+1)[2\hbar^2\alpha^2(n+1)+\omega^2]\}$ \\[0.2cm]
  DRHO & $-4\lambda a^2$ & $2n\hbar\omega - 4\hbar\lambda n(l+1+\hbar n)$ \\[0.2cm]
  DC & $-\lambda a^2 - \frac{e^4}{4a^2}$ & $n\hbar(2l+2+n\hbar)$ \\[0.2cm]
  & & $ \times \left(-\lambda+\frac{e^4}{4(l+1)^2(l+1+n\hbar)^2}\right)$ \\[0.2cm]
  \hline 
\end{tabular}
\end{center}

\end{table}
\par
%
%
\section{\boldmath Deformed Shape Invariance for Superpotentials with an Explicit Dependence on $\hbar$}

\setcounter{equation}{0}

Let us next consider the case where the superpotential contains an explicit dependence on $\hbar$. It may then be expanded in powers of $\hbar$ as
\begin{equation}
  W(x,a,\hbar) = \sum_{n=0}^{\infty} \hbar^n W_n(x,a).
\end{equation}
On inserting this expression in the DSI condition (\ref{eq:DSI}) and proceeding as in conventional SUSYQM \cite{bougie10, bougie12}, we arrive at the set of relations
\begin{align}
  & \sum_{k=0}^n W_k W_{n-k} + f \frac{\partial W_{n-1}}{\partial x} - \sum_{s=0}^n \sum_{k=0}^s
       \frac{1}{(n-s)!} \frac{\partial^{n-s}}{\partial a^{n-s}} W_k W_{s-k} \nonumber \\
  & {}+ f \sum_{k=1}^n \frac{1}{(k-1)!} \frac{\partial^k}{\partial a^{k-1}\partial x} W_{n-k} - \frac{1}{n!}
       \frac{d^ng}{da^n} = 0, \qquad n=1, 2, \ldots.
\end{align}
The latter can be rewritten as
\begin{align}
  & 2f \frac{\partial W_0}{\partial x} - \frac{\partial}{\partial a} (W_0^2 + g) = 0,  \label{eq:equation1} \\
  & f \frac{\partial W_1}{\partial x} - \frac{\partial}{\partial a}(W_0W_1) = 0,  \label{eq:equation2} \\
  & 2f \frac{\partial W_{n-1}}{\partial x} - \sum_{s=1}^{n-1} \sum_{k=0}^s \frac{1}{(n-s)!} \frac{\partial^{n-s}}
       {\partial a^{n-s}} W_k W_{s-k} + \frac{n-2}{n!} f \frac{\partial^n W_0}{\partial a^{n-1}\partial x} \nonumber \\
  & {}+ f \sum_{k=2}^{n-1} \frac{1}{(k-1)!} \frac{\partial^k}{\partial a^{k-1} \partial x} W_{n-k} = 0,
       \qquad n=3, 4, \ldots.  \label{eq:equation3}
\end{align}
\par
%
%
In \cite{cq16}, two sets of rational extensions of the DRHO potential with $\lambda<0$ considered in Section 3, referred to as type I and type II extensions, were constructed in terms of some Jacobi polynomials of degree $m$. The potentials belonging to these two sets were shown to be derived from superpotentials satisfying the DSI condition. The simplest potentials, corresponding to $m=1$, turn out to be identical and given by (after changing the parameters and the variable as explained in the appendix)
\begin{align}
  V &= \frac{\omega(\omega-2\hbar|\lambda|)x^2}{4(1-|\lambda|x^2)} + \frac{(l+1)(l+1-\hbar)}{x^2}
       + 4\hbar^2 \biggl(\frac{\omega+2|\lambda|(l+1-\hbar)}{[\omega-2|\lambda|(l+1)]x^2 + 2l+2-\hbar}
       \nonumber \\
  & \quad {}- \frac{2(2l+2-\hbar)(\omega-\hbar|\lambda|)}{\{[\omega-2|\lambda|(l+1)]x^2+2l+2-\hbar\}^2}
       \biggr),
\end{align}
with a corresponding superpotential
\begin{align}
  W &= \frac{\omega x}{2\sqrt{1-|\lambda|x^2}} - \frac{l+1}{x}\sqrt{1-|\lambda|x^2} + 2\hbar [\omega
       -2(l+1)|\lambda|] x \sqrt{1-|\lambda|x^2} \nonumber \\
  & \quad {}\times \biggl(\frac{1}{[\omega-2(l+1)|\lambda|]x^2+2l+2-\hbar} -
       \frac{1}{[\omega-2(l+1)|\lambda|]x^2+2l+2+\hbar}\biggr).  \label{eq:W} 
\end{align}.
Let us show that such a superpotential can be derived by the present method.\par
%
%
{}For such a purpose, we plan to prove that for
\begin{equation}
  f = \sqrt{1-|\lambda|x^2}, \qquad a = \frac{1}{2}\left(l+1+\frac{\omega}{2|\lambda|}\right), \qquad
  b = \frac{1}{2}\left(l+1-\frac{\omega}{2|\lambda|}\right),  \label{eq:definitions} 
\end{equation}
the functions
\begin{align}
  & W_0(x,a) = - \frac{a+b}{x}f + \frac{(a-b)|\lambda|x}{f}, \label{eq:W0} \\
  & W_{2\nu+1}(x,a) = 0, \qquad \nu=0, 1, 2, \ldots, \\
  & W_{2\nu}(x,a) = - f \frac{16b|\lambda|x}{(4bf^2+2a-2b)^{2\nu}}, \qquad \nu=1, 2, \ldots,
     \label{eq:W2nu}
\end{align}
provide a solution of the set of equations (\ref{eq:equation1}), (\ref{eq:equation2}), and (\ref{eq:equation3}). Note that, as in Section~3, the combinations of parameters $a$ and $b$ become $a+\hbar$ and $b$ for the partner, respectively.\par
%
%
Let us start with Eq.~(\ref{eq:equation1}). From (\ref{eq:W0}), we get
\begin{align}
  \frac{\partial W_0}{\partial x} &= \frac{a+b}{x^2}f + \frac{(a+b)|\lambda|}{f} + \frac{(a-b)|\lambda|}{f^3}, \\
  \frac{\partial W_0}{\partial a} &= - \frac{1}{x}f + \frac{|\lambda|x}{f},
\end{align}
from which we obtain $\frac{dg}{da} = 8a|\lambda|$ and $g=4a^2|\lambda|$ up to some additive constant. Hence, from Eq.~(\ref{eq:energy}),
\begin{equation}
  E^{(-)}_n = 4\hbar |\lambda|n(n\hbar+2a) = 4\hbar |\lambda|n \left(n\hbar +l+1+\frac{\omega}{2|\lambda|}
  \right),
\end{equation}
in agreement with the result obtained in \cite{cq16}.\par
%
%
Equation~(\ref{eq:equation2}) is automatically satisfied since $W_1=0$.\par
%
%
Considering next Eq.~(\ref{eq:equation3}), we note that
\begin{equation}
  \frac{\partial^n W_0}{\partial a^{n-1}\partial x} = 0, \qquad n=3, 4, \ldots,  \label{eq:result1}
\end{equation}
and that
\begin{equation}
  \sum_{k=0}^s W_k W_{s-k} = 0 \qquad \text{for odd $s$}.  \label{eq:odd-s}
\end{equation}
For even $s$, on the other hand, we easily get
\begin{equation}
  \sum_{k=0}^s W_k W_{s-k} = F_s,
\end{equation}
with $F_s$ defined by
\begin{equation}
  F_s = \frac{32b|\lambda|}{(4bf^2+2a-2b)^s} \{-4b(s-2)f^4 + [2a+4b(s-2)]f^2 - a +b\}.
\end{equation}
From this result, it is straightforward to prove that 
\begin{equation}
  \frac{\partial^{n-s}}{\partial a^{n-s}} \sum_{k=0}^s W_k W_{s-k} = (-2)^{n-s} \frac{(n-2)!}{(s-2)!} F_n
  \qquad \text{for even $s$}.  \label{eq:even-s} 
\end{equation}
Equations (\ref{eq:odd-s}) and (\ref{eq:even-s}) then lead to
\begin{equation}
  \sum_{s=1}^{n-1} \sum_{k=0}^s \frac{1}{(n-s)!} \frac{\partial^{n-s}}{\partial a^{n-s}} W_k W_{s-k} =
  F_n \times \begin{cases}
     \frac{1}{2}(3^{n-2}-1) & \text{for even $n$}, \\
     - \frac{1}{2}(3^{n-2}+1) & \text{for odd $n$}.  
  \end{cases}. \label{eq:result2}
\end{equation}
\par
%
%
{}Furthermore, we obtain
\begin{equation}
  2f \frac{\partial W_{n-1}}{\partial x} = \begin{cases}
      0 & \text{for even $n$}, \\
      - 2F_n & \text{for odd $n$},
  \end{cases}. \label{eq:result3}
\end{equation}
as well as
\begin{equation}
  f \sum_{k=2}^{n-1} \frac{1}{(k-1)!} \frac{\partial^k}{\partial a^{k-1}\partial x} W_{n-k} = F_n \times
  \begin{cases}
     \frac{1}{2}(3^{n-2}-1) & \text{for even $n$}, \\
     - \frac{1}{2}(3^{n-2}-3) & \text{for odd $n$}.
  \end{cases}. \label{eq:result4}
\end{equation}
\par
%
%
On inserting Eqs.~(\ref{eq:result1}), (\ref{eq:result2}), (\ref{eq:result3}), and (\ref{eq:result4}) in Eq.~(\ref{eq:equation3}), it is clear that the latter is satisfied, which completes the proof that Eqs.~(\ref{eq:W0})--(\ref{eq:W2nu}) provide a solution of Eqs.~(\ref{eq:equation1})--(\ref{eq:equation3}).\par
%
%
It now only remains to use Eqs.~(\ref{eq:W0}) and (\ref{eq:W2nu}) in
\begin{equation}
  W(x,a,\hbar) = \sum_{\nu=0}^{\infty} \hbar^{2\nu} W_{2\nu}(x,a)
\end{equation}
to obtain
\begin{align}
  W(x,a,\hbar) &= \frac{(a-b)|\lambda|x}{f} - \frac{a+b}{x}f - 8\hbar b|\lambda|xf \biggl(\frac{1}{4bf^2+2a
      -2b-\hbar} \nonumber \\
  & \quad {}- \frac{1}{4bf^2+2a-2b+\hbar}\biggr),
\end{align}
which, after introducing the definitions of $f$, $a$, and $b$, given in (\ref{eq:definitions}), reduces to the expression (\ref{eq:W}), i.e., the extended superpotential found in \cite{cq16}.\par
%
%
\section{Conclusion}

In this paper, we have shown that the approach of Gangopadhyaya, Mallow, and their coworkers of SI potentials in conventional SUSYQM can be extended to DSI ones in DSUSYQM and we have illustrated our results by considering several examples taken from \cite{bagchi05, cq09b, cq16}. These include both conventional potentials, for which the corresponding superpotential has no explicit dependence on $\hbar$, and a rationally-extended one, for which there is such a dependence. In all cases, it turns out that the parameter $a$, which is changed into $a+\hbar$ when going to the partner potential, is a combination of the potential and deforming function parameters.\par
%
%
An interesting open question for future investigation would be the possibility of generalizing the method to rationally-extended potentials exhibiting an ``enlarged'' shape invariance, for which the partner is obtained by translating some potential parameter as well as the degree $m$ of the polynomial arising in the denominator. Such potentials are indeed known both in conventional SUSYQM \cite{cq12a, cq12b, grandati12, grandati15} and in DSUSYQM \cite{cq16}.\par
%
%
\section*{Acknowledgments}

This work was supported by the Fonds de la Recherche Scientifique - FNRS under Grant Number 4.45.10.08.\par
%
%
\section*{Appendix: Going from Previously Used Conventions to the Present Ones}

\renewcommand{\theequation}{A.\arabic{equation}}
\setcounter{equation}{0}

In this appendix, we summarize the changes that have to be carried out to go from the conventions used in \cite{bagchi05, cq09b, cq16} to those of the present paper. The quantities employed in the former papers are distinguished by a bar from those used here. It is worth noting too that in \cite{bagchi05, cq09b, cq16}, the potentials used have a nonvanishing ground-state energy and must therefore be compared with $V = V_- + E_0$, where $E_0$ is the shift to adjust the ground-state energy of $H_-$ to a zero value. As a consequence,  $E^{(-)}_n = E_n - E_0$ corresponds to $\bar{E}_n - \bar{E}_0$.\par
%
%
\noindent
{\it P\"oschl-Teller potential}
\begin{equation}
\begin{split}
  &\bar{V} = \bar{A}(\bar{A}-1) \sec^2 x, \quad \bar{f}(\bar{x}) = 1 + \bar{\alpha}\sin^2 \bar{x}, \\
  &\bar{A} = \frac{A}{\hbar}, \quad \bar{\alpha} = \alpha, \quad \bar{x} = x, \\
  &V = A(A-\hbar)\sec^2 x = \hbar^2 \bar{V}, \quad f(x) = 1 + \alpha \sin^2 x = \bar{f}(\bar{x}), \\
  &W = \hbar \bar{W}, \quad E_n = \hbar^2 \bar{E}_n. 
\end{split}
\end{equation}
%
%
\noindent
{\it Radial harmonic oscillator potential}
\begin{equation}
\begin{split}
  &\bar{V} = \frac{1}{4}\bar{\omega}^2\bar{x}^2 + \frac{\bar{l}(\bar{l}+1)}{\bar{x}^2}, \quad \bar{f}(\bar{x})
       = 1 + \bar{\alpha}\bar{x}^2, \\
  &\bar{\omega} = \hbar\omega, \quad \bar{l}=\frac{l+1}{\hbar}-1, \quad  \bar{\alpha} = \hbar^2\alpha, \quad 
       \bar{x} = \frac{x}{\hbar}, \\
  &V = \frac{1}{4}\omega^2x^2 + \frac{(l+1)(l+1-\hbar)}{x^2} = \bar{V}, \quad f(x) = 1 + \alpha x^2 = 
       \bar{f}(\bar{x}), \\
  &W = \bar{W}, \quad E_n = \bar{E}_n. 
\end{split}
\end{equation}
%
%
\noindent
{\it Scarf I potential}
\begin{equation}
\begin{split}
  &\bar{V} = [\bar{A}(\bar{A}-1)]+\bar{B}^2] \sec^2\bar{x} - \bar{B}(2\bar{A}-1)\sec\bar{x}\tan\bar{x},
      \quad \bar{f}(\bar{x}) = 1 + \bar{\alpha}\sin\bar{x}, \\
  &\bar{A} = \frac{A}{\hbar}, \quad \bar{B}=\frac{B}{\hbar}, \quad \bar{\alpha} = \alpha, \quad \bar{x} = x, \\
  &V = [A(A-\hbar)+B^2]\sec^2x - B(2A-\hbar)\sec x\tan x = \hbar^2\bar{V}, \\
  &f(x) = 1+\alpha\sin x = \bar{f}(\bar{x}), \quad W = \hbar\bar{W}, \quad  E_n = \hbar^2\bar{E}_n. 
\end{split}
\end{equation}
%
%
\noindent
{\it Coulomb potential}
\begin{equation}
\begin{split}
  &\bar{V} = - \frac{2\bar{Z}}{\bar{x}} + \frac{\bar{l}(\bar{l}+1)}{\bar{x}^2}, \quad \bar{f}(\bar{x})
       = 1 + \bar{\alpha}\bar{x}, \\
  &\bar{Z} = \frac{e^2}{2\hbar}, \quad \bar{l}=\frac{l+1}{\hbar}-1, \quad  \bar{\alpha} = \hbar\alpha, \quad 
       \bar{x} = \frac{x}{\hbar}, \\
  &V = -\frac{e^2}{x} + \frac{(l+1)(l+1-\hbar)}{x^2} = \bar{V}, \quad f(x) = 1 + \alpha x = 
       \bar{f}(\bar{x}), \\
  &W = \bar{W}, \quad E_n = \bar{E}_n. 
\end{split}
\end{equation}
%
%
\noindent
{\it Morse potential}
\begin{equation}
\begin{split}
  &\bar{V} = \bar{B}^2 e^{-2\bar{x}} - \bar{B}(2\bar{A}+1)e^{-x}, \quad \bar{f}(\bar{x})
       = 1 + \bar{\alpha}e^{-\bar{x}}, \\
  &\bar{A} = \frac{A}{\hbar}, \quad \bar{B}=\frac{B}{\hbar}, \quad  \bar{\alpha} = \alpha, \quad 
       \bar{x} = x, \\
  &V = B^2 e^{-2x} - B(2A+\hbar)e^{-x} = \hbar^2\bar{V}, \quad f(x) = 1 + \alpha e^{-x} = 
       \bar{f}(\bar{x}), \\
  &W = \hbar\bar{W}, \quad E_n = \hbar^2\bar{E}_n. 
\end{split}
\end{equation}
%
%
\noindent
{\it Eckart potential}
\begin{equation}
\begin{split}
  &\bar{V} = \bar{A}(\bar{A}-1)\csch^2\bar{x} - 2\bar{B}\coth\bar{x}, \quad \bar{f}(\bar{x})
       = 1 + \bar{\alpha}e^{-\bar{x}}\sinh\bar{x}, \\
  &\bar{A} = \frac{A}{\hbar}, \quad \bar{B}=\frac{B}{\hbar^2}, \quad  \bar{\alpha} = \alpha, \quad 
       \bar{x} = x, \\
  &V = A(A-\hbar)\csch^2x - 2B\coth x = \hbar^2\bar{V}, \quad f(x) = 1 + \alpha e^{-x}\sinh x = 
       \bar{f}(\bar{x}), \\
  &W = \hbar\bar{W}, \quad E_n = \hbar^2\bar{E}_n. 
\end{split}
\end{equation}
%
%
\noindent
{\it Rosen-Morse I potential}
\begin{equation}
\begin{split}
  &\bar{V} = \bar{A}(\bar{A}-1)\csc^2\bar{x} + 2\bar{B}\cot\bar{x}, \quad \bar{f}(\bar{x})
       = 1 + \sin\bar{x}(\bar{\alpha}\cos\bar{x}+\bar{\beta}\sin\bar{x}), \\
  &\bar{A} = \frac{A}{\hbar}, \quad \bar{B}=\frac{B}{\hbar^2}, \quad  \bar{\alpha} = \alpha, \quad 
       \bar{\beta} = \beta, \quad \bar{x} = x, \\
  &V = A(A-\hbar)\csc^2x + 2B\cot x = \hbar^2\bar{V}, \quad f(x) = 1 + \sin x(\alpha \cos x + \beta\sin x) = 
       \bar{f}(\bar{x}), \\
  &W = \hbar\bar{W}, \quad E_n = \hbar^2\bar{E}_n. 
\end{split}
\end{equation}
%
%
\noindent
{\it Shifted harmonic oscillator potential}
\begin{equation}
\begin{split}
  &\bar{V} = \frac{1}{4}\bar{\omega}^2\left(\bar{x}-\frac{2\bar{d}}{\bar{\omega}}\right)^2, \quad 
       \bar{f}(\bar{x}) = 1 + \bar{\alpha}\bar{x}^2 + 2\bar{\beta}\bar{x}, \\
  &\bar{\omega} = \hbar\omega, \quad \bar{d}=d, \quad  \bar{\alpha} = \hbar^2\alpha, \quad 
       \bar{\beta} = \hbar\beta, \quad \bar{x} = \frac{x}{\hbar}, \\
  &V = \frac{1}{4}\omega^2\left(x-\frac{2d}{\omega}\right)^2 = \bar{V}, \quad f(x) = 1 + \alpha x^2 +        
       2\beta x = \bar{f}(\bar{x}), \\
  &W = \bar{W}, \quad E_n = \bar{E}_n. 
\end{split}
\end{equation}
%
%
\noindent
{\it Deformed radial harmonic oscillator potential}
\begin{equation}
\begin{split}
  &\bar{V} = \frac{\bar{\omega}(\bar{\omega}+2\bar{\lambda})\bar{x}^2}{4(1+\bar{\lambda}\bar{x}^2)} +
       \frac{\bar{l}(\bar{l}+1)}{\bar{x}^2}, \quad \bar{f}(\bar{x}) = \sqrt{1 + \bar{\lambda}\bar{x}^2}, \\
  &\bar{\omega} = \hbar\omega, \quad \bar{l}=\frac{l+1}{\hbar}-1, \quad  \bar{\lambda} = \hbar^2\lambda, 
       \quad \bar{x} = \frac{x}{\hbar}, \\
  &V = \frac{\omega(\omega+2\hbar\lambda)x^2}{4(1+\lambda x^2)} + \frac{(l+1)(l+1-\hbar)}{x^2} = 
       \bar{V}, \quad f(x) = \sqrt{1+\lambda x^2} = \bar{f}(\bar{x}), \\
  &W = \bar{W}, \quad E_n = \bar{E}_n. 
\end{split}
\end{equation}
%
%
\noindent
{\it Deformed Coulomb potential}
\begin{equation}
\begin{split}
  &\bar{V} = - \frac{\bar{Q}}{\bar{x}} \sqrt{1+\bar{\lambda}\bar{x}^2} +
       \frac{\bar{l}(\bar{l}+1)}{\bar{x}^2}, \quad \bar{f}(\bar{x}) = \sqrt{1 + \bar{\lambda}\bar{x}^2}, \\
  &\bar{Q} = \frac{e^2}{\hbar}, \quad \bar{l}=\frac{l+1}{\hbar}-1, \quad  \bar{\lambda} = \hbar^2\lambda, 
       \quad \bar{x} = \frac{x}{\hbar}, \\
  &V = - \frac{e^2}{x} \sqrt{1+\lambda x^2} + \frac{(l+1)(l+1-\hbar)}{x^2} = 
       \bar{V}, \quad f(x) = \sqrt{1+\lambda x^2} = \bar{f}(\bar{x}), \\
  &W = \bar{W}, \quad E_n = \bar{E}_n. 
\end{split}
\end{equation}
%
%


\newpage

\begin{thebibliography}{99}

\bibitem{cooper}
Cooper, F.; Khare, A.; Sukhatme, U.
Supersymmetry and quantum mechanics.
{\it Phys.\ Rep.} {\bf 1995}, {\it 251}, 267--385.

\bibitem{junker}
Junker, G.
{\it Supersymmetric Methods in Quantum and Statistical Physics};
Springer-Verlag: Berlin, 1996.

\bibitem{bagchi00}
Bagchi, B.
{\it Supersymmetry in Quantum and Classical Physics};
Chapman and Hall/CRC: Boca Raton, FL, 2000.

\bibitem{gango10}
Gangopadhyaya, A.; Mallow, J.; Rasinariu, C.
{\it Supersymmetric Quantum Mechanics: An Introduction};
World Scientific: Singapore, 2010.

\bibitem{genden}
Gendenshtein, L.
Derivation of exact spectra of the Schr\"odinger equation by means of supersymmetry.
{\it JETP Lett.} {\bf 1983}, {\it 38}, 356--359.

\bibitem{darboux}
Darboux, G.
{\it Le\c cons sur la Th\'eorie G\'en\'erale des Surfaces}, 2nd ed.;
Gauthier-Villars: Paris, 1912.

\bibitem{schro40}
Schr\"odinger, E.
A method of determining quantum-mechanical eigenvalues and eigenfunctions.
{\it Proc.\ R.\ Ir.\ Acad.} {\bf 1940}, {\it A46}, 9--16.

\bibitem{schro41a}
Schr\"odinger, E.
Further studies on solving eigenvalue problems by factorization.
{\it Proc.\ R.\ Ir.\ Acad.} {\bf 1941}, {\it A46}, 183--206.

\bibitem{schro41b}
Schr\"odinger, E.
The factorization of the hypergeometric equation.
{\it Proc.\ R.\ Ir.\ Acad.} {\bf 1941}, {\it A47}, 53--54.

\bibitem{infeld51}
Infeld, L.; Hull, T.E.
The factorization method.
{\it Rev.\ Mod.\ Phys.} {\bf 1951}, {\it 23}, 21--68.

\bibitem{gomez14}
G\'omez-Ullate, D.; Grandati, Y.; Milson, R.
Extended Krein-Adler theorem for the translationally shape invariant potentials.
{\it J.\ Math.\ Phys.} {\bf 2014}, {\it 55}, 043510, 30 pages.

\bibitem{cq08}
Quesne, C.
Exceptional orthogonal polynomials, exactly solvable potentials and supersymmetry.
{\it J.\ Phys.\ A} {\bf 2008}, {\it 41}, 392001, 6 pages.

\bibitem{bagchi09}
Bagchi, B.; Quesne, C.; Roychoudhury, R.
Isospectrality of conventional and new extended potentials, second-order supersymmetry and role of $\cal PT$ symmetry.
{\it Pramana J.\ Phys.} {\bf 2009}, {\it 73}, 337--347.

\bibitem{cq09a}
Quesne, C.
Solvable rational potentials and exceptional orthogonal polynomials in supersymmetric quantum mechanics.
{\it SIGMA} {\bf 2009}, {\it 5}, 084, 24 pages.

\bibitem{odake}
Odake, S.; Sasaki, R.
Infinitely many shape invariant potentials and new orthogonal polynomials.
{\it Phys.\ Lett.\ B} {\bf 2009}, {\it 679}, 414--417.

\bibitem{gomez09}
G\'omez-Ullate, D.; Kamran, R.; Milson, R.
An extended class of orthogonal polynomials defined by a Sturm-Liouville problem.
{\it J.\ Math.\ Anal.\ Appl.} {\bf 2009}, {\it 359}, 352--367.

\bibitem{kempf94}
Kempf, A.
Uncertainty relation in quantum mechanics with quantum group symmetry.
{\it J.\ Math.\ Phys.} {\bf 1994}, {\it 35}, 4483--4496.

\bibitem{hinrichsen}
Hinrichsen, H.; Kempf, A.
Maximal localization in the presence of minimal uncertainties in positions and in momenta.
{\it J.\ Math.\ Phys.} {\bf 1996}, {\it 37}, 2121--2137.

\bibitem{kempf97}
Kempf, A.
Non-pointlike particles in harmonic oscillators.
{\it J.\ Phys.\ A} {\bf 1997}, {\it 30}, 2093--2102.

\bibitem{witten}
Witten, E.
Reflections on the fate of spacetime.
{\it Phys.\ Today} {\bf 1996}, {\it 49}, 24--30.

\bibitem{bastard}
Bastard, G.
{\it Wave Mechanics Applied to Semiconductor Heterostructures};
Editions de Physique: Les Ulis, 1988.

\bibitem{weisbuch}
Weisbuch, C; Vinter, B.
{\it Quantum Semiconductor Heterostructures};
Academic: New York, 1997.

\bibitem{serra}
Serra, L.; Lipparini, E.
Spin response of unpolarized quantum dots.
{\it Europhys.\ Lett.} {\bf 1997}, {\it 40}, 667--672.

\bibitem{harrison}
Harrison, P.;Valavanis, A.
{\it Quantum Wells, Wires and Dots: Theoretical and Computational Physics of Semiconductor Nanostructures};
Wiley: Chichester, 2016.

\bibitem{barranco}
Barranco, M.; Pi, M.; Gatica, S.M.; Hern\'andez, E.S.; Navarro, J.
Structure and energetics of mixed $^4$He-$^3$He drops.
{\it Phys.\ Rev.\ B} {\bf 1997}, {\it 56}, 8997--9003.

\bibitem{geller}
Geller, M.R.; Kohn, W.
Quantum mechanics of electrons in crystals with graded composition.
{\it Phys.\ Rev.\ Lett.} {\bf 1993}, {\it 70}, 3103--3106.

\bibitem{arias}
Arias de Saavedra, F.; Boronat, J.; Polls, A.; Fabrocini, A.
Effective mass of one $^4$He atom in liquid $^3$He.
{\it Phys.\ Rev. B} {\bf1994}, {\it 50}, 4248(R)--4251(R).

\bibitem{puente}
Puente, A.; Serra, Ll.; Casas, M.
Dipole excitation of Na clusters with a non-local energy density functional.
{\it Z.\ Phys.\ D} {\bf 1994}, {\it 31}, 283--286.

\bibitem{ring}
Ring, P.; Schuck, P.
{\it The Nuclear Many Body Problem};
Springer: New York, 1980.

\bibitem{bonatsos}
Bonatsos, D.; Georgoudis, P.E.; Lenis, D.; Minkov, N.; Quesne, C.
Bohr Hamiltonian with a deformation-dependent mass term for the Davidson potential.
{\it Phys.\ Rev.\ C} {\bf 2011}, {\it 83}, 044321, 20 pages.

\bibitem{willatzen}
Willatzen, M.; Lassen B.
The Ben Daniel - Duke model in general nanowire structures.
{\it J.\ Phys.:\ Condens.\ Matter} {\bf 2007}, {\it 19}, 136217, 8 pages.

\bibitem{chamel}
Chamel, N.
Effective mass of free neutrons in neutron star crust.
{\it Nucl.\ Phys.\ A} {\bf 2006}, {\it 773}, 263--278.

\bibitem{infeld41}
Infeld, L.
On a new treatment of some eigenvalue problems.
{\it Phys.\ Rev.} {\bf 1941}, {\it 59}, 737--747.

\bibitem{stevenson}
Stevenson, A.F.
Note on the ``Kepler problem'' in a spherical space, and the factorization method of solving eigenvalue problems.
{\it Phys.\ Rev.} {\bf 1941}, {\it 59}, 842.

\bibitem{infeld45}
Infeld, L.; Schild, A.
A note on the Kepler problem in a space of constant negative curvature.
{\it Phys.\ Rev.} {\bf 1945}, {\it 67}, 121.

\bibitem{kalnins96}
Kalnins, E.G.; Miller, W.\ Jr; Pogosyan, G.S.
Superintegrability and associated polynomial solutions: Euclidean space and the sphere in two dimensions.
{\it J.\ Math.\ Phys.} {\bf 1996}, {\it 37}, 6439--6467.

\bibitem{kalnins97}
Kalnins, E.G.; Miller, W.\ Jr; Pogosyan, G.S.
Superintegrability on the two-dimensional hyperboloid.
{\it J.\ Math.\ Phys.} {\bf 1997}, {\it 38}, 5416--5433.

\bibitem{cq04}
Quesne, C.; Tkachuk, V.M.
Deformed algebras, position-dependent effective masses and curved spaces: an exactly solvable Coulomb problem.
{\it J.\ Phys.\ A} {\bf 2004}, {\it 37}, 4267--4281.

\bibitem{bagchi05}
Bagchi, B.; Banerjee, A.; Quesne, C.; Tkachuk, V.M.
Deformed shape invariance and exactly solvable Hamiltonians with position-dependent effective mass.
{\it J.\ Phys.\ A} {\bf 2005}, {\it 38}, 2929--2945.

\bibitem{cq09b}
Quesne, C.
Point canonical transformations versus deformed shape invariance for position-dependent mass Sch\"odinger equations.
{\it SIGMA} {\bf 2009}, {\it 5}, 046, 17 pages.

\bibitem{cq16}
Quesne, C.
Quantum oscillator and Kepler-Coulomb problems in curved spaces: Deformed shape invariance, point canonical transformations, and rational extensions.
{\it J.\ Math.\ Phys.} {\bf 2016}, {\it 57}, 102101, 20 pages.

\bibitem{gango08}
Gangopadhyaya, A.; Mallow, J.V.
Generating shape invariant potentials.
{\it Int.\ J.\ Mod.\ Phys.\ A} {\bf 2008}, {\it 23}, 4949--4978.

\bibitem{bougie10}
Bougie, J.; Gangopadhyaya, A.; Mallow, J.V.
Generation of a complete set of additive shape-invariant potentials from an Euler equation.
{\it Phys.\ Rev.\ Lett.} {\bf 2010} {\it 105}, 210402, 4 pages. 

\bibitem{bougie12}
Bougie, J.; Gangopadhyaya, A.; Mallow, J.; Rasinariu, C.
Supersymmetric quantum mechanics and solvable models.
{\it Symmetry} {\bf 2012}, {\it 4}, 452--473.

\bibitem{bougie15}
Bougie, J.; Gangopadhyaya, A.; Mallow, J.V; Rasinariu, C.
Generation of a novel exactly solvable potential.
{\it Phys.\ Lett.\ A} {\bf 2015}, {\it 379}, 2180--2183.

\bibitem{mallow}
Mallow, J.V.; Gangopadhyaya, A.; Bougie, J.; Rasinariu, C.
Inter-relations between additive shape invariant superpotentials.
{\it Phys.\ Lett.\ A} {\bf 2020}, {\it 384}, 126129, 6 pages.

\bibitem{mustafa}
Mustafa, O.; Mazharimousavi, S.H.
Ordering ambiguity revisited via position dependent mass pseudo-momentum operators.
{\it Int.\ J.\ Theor.\ Phys.} {\bf 2007}, {\it 46}, 1786--1796.

\bibitem{cq12a}
Quesne, C.
Revisiting (quasi-)exactly solvable rational extensions of the Morse potential.
{\it Int.\ J.\ Mod.\ Phys.\ A} {\bf 2012}, {\it 27}, 1250073, 18 pages.

\bibitem{cq12b}
Quesne, C.
Novel enlarged shape invariance property and exactly solvable rational extensions of the Rosen-Morse II and Eckart potentials.
{\it SIGMA} {\bf 2012}, {\it 8}, 080, 19 pages.

\bibitem{grandati12}
Grandati, Y.
New rational extensions of solvable potentials with finite bound state spectrum.
{\it Phys.\ Lett.\ A} {\bf 2012}, {\it 376}, 2866--2872.

\bibitem{grandati15}
Grandati, Y.; Quesne, C.
Confluent chains of DBT: Enlarged shape invariance and new orthogonal polynomials.
{\it SIGMA} {\bf 2015}, {\it 11}, 061, 26 pages.

\end{thebibliography}
\end{document}